\documentclass[aps,twocolumn,nofootinbib,showkeys,showpacs,superscriptaddress]{revtex4-1}

\usepackage{amssymb}
\usepackage{amsmath}
\usepackage{amstext}
\usepackage{amsfonts}
\usepackage{bbold}
\usepackage{bm}
\usepackage{graphics}
\usepackage{mathtools}
\usepackage{lmodern}
\usepackage{graphicx}
\usepackage{hyperref}
\usepackage{xcolor}
\usepackage[export]{adjustbox}
\usepackage{soul}
\usepackage{microtype}
\usepackage{xspace}
\usepackage{array}

\definecolor{myred}{RGB}{220, 0, 0}

\newcommand{\GeV}{\ensuremath{{\mathrm{\,Ge\kern -0.1em V}}}\xspace}
\newcommand{\MeV}{\ensuremath{{\mathrm{\,Me\kern -0.1em V}}}\xspace}
\newcommand{\keV}{\ensuremath{{\mathrm{\,ke\kern -0.1em V}}}\xspace}
\newcommand{\eV}{\ensuremath{{\mathrm{\,e\kern -0.1em V}}}\xspace}
\newcommand{\meV}{\ensuremath{{\mathrm{\,me\kern -0.1em V}}}\xspace}

\newcommand{\del}{\bm\nabla}
\newcommand{\mHe}{m_\text{He}}
\newcommand{\cs}{c_s}
\newcommand{\nbar}{\bar n}
\newcommand{\sprod}[2]{\bm{q}_{#1}\cdot\bm{q}_{#2}}
\newcommand{\He}{$^4$He\xspace}
\newcommand{\mchi}{m_\chi}

\newcolumntype{C}{>{$}c<{$}}

\begin{document}

\title{Directional detection of light dark matter from three-phonon events in superfluid \He} 

\author{Andrea~Caputo}
\affiliation{School of Physics and Astronomy, Tel-Aviv University, Tel-Aviv 69978, Israel}
\affiliation{Department of Particle Physics and Astrophysics,
Weizmann Institute of Science, Rehovot 7610001, Israel}
\affiliation{Max-Planck-Institut f\"ur Physik (Werner-Heisenberg-Institut), F\"ohringer Ring 6, 80805 M\"unchen, Germany}

\author{Angelo~Esposito}
\affiliation{Theoretical Particle Physics Laboratory (LPTP), Institute of Physics, EPFL, 1015 Lausanne, Switzerland}

\author{Fulvio~Piccinini}
\affiliation{INFN, Sezione di Pavia, via A. Bassi 6, 27100 Pavia, Italy}

\author{Antonio~D.~Polosa}
\affiliation{Dipartimento di Fisica, Sapienza Universit\`a di Roma, Piazzale Aldo Moro 2, 00185 Roma, Italy}
\affiliation{INFN Sezione di Roma, Piazzale Aldo Moro 2, 00185 Roma, Italy}

\author{Giuseppe~Rossi}
\affiliation{Dipartimento di Fisica, Sapienza Universit\`a di Roma, Piazzale Aldo Moro 2, 00185 Roma, Italy}

\begin{abstract}
We present the analysis of a new signature for light dark matter detection with superfluid \He: the emission of three phonons. We show that, in a region of mass below the MeV, the kinematics of this process can offer a way to reconstruct the dark matter interaction vertex, while providing  background rejection via coincidence requirements and directionality. We develop all the {theoretical tools} to deal with such an observable, and compute the associated differential distributions.
\end{abstract}


\maketitle


\section{Introduction}

While there is overwhelming evidence that the largest fraction of matter in the Universe is dark matter, little is known about its nature. 
In particular, the possible dark matter mass spans several orders of magnitude, and different masses require vastly different detection techniques. Recently, increasing attention has been paid to candidates in the keV to GeV mass range (see~\cite{
Boehm:2003ha,Boehm:2003hm,Hooper:2008im,Feng:2008ya,Hall:2009bx,Falkowski:2011xh,Hochberg:2014dra,DAgnolo:2015ujb,Hochberg:2015vrg,Kuflik:2015isi,Green:2017ybv,DAgnolo:2018wcn,Mondino:2020lsc}, and~\cite{Battaglieri:2017aum,Knapen:2017xzo} for a review), which, while massive enough to be treated as pointlike, cannot release appreciable energy to a material via standard recoil processes, hence requiring detectors with low energy thresholds. When the typical exchanged momentum is below the keV, the prime signature of the interaction with such particles is the emission of collective excitations in the detector material. Several proposals have been put forth along these lines~\cite{Hochberg:2015pha,Hochberg:2016ajh,Guo:2013dt,Schutz:2016tid,Knapen:2016cue,Hertel:2018aal,Acanfora:2019con,Caputo:2019cyg,Caputo:2019xum,Baym:2020uos,Knapen:2017ekk,Campbell-Deem:2019hdx,Griffin:2018bjn,Griffin:2019mvc,Trickle:2019nya,Hochberg:2017wce,Geilhufe:2019ndy,Coskuner:2019odd,Trickle:2019ovy,Arvanitaki:2017nhi,Lawson:2019brd, Gelmini:2020xir, Gelmini:2020kcu,Bunting:2017net,Chen:2020jia,Cox:2019cod,Capparelli:2014lua,Cavoto:2017otc,Trickle:2020oki,Dror:2020czw, Griffin:2020lgd}.

A promising direction is that of considering the emission of collective excitations in superfluid \He~\cite{Guo:2013dt,Schutz:2016tid,Knapen:2016cue,Hertel:2018aal,Acanfora:2019con,Caputo:2019cyg,Caputo:2019xum,Baym:2020uos}, in particular gapless phonons for small dark matter masses. It has been shown that the process where the dark matter interacts with the bulk of the detector and emits two excitations has a favorable kinematics, allowing to probe masses down to the warm dark matter limit, $\mchi\sim\keV$~\cite{Schutz:2016tid,Knapen:2016cue}. 

In~\cite{Acanfora:2019con,Caputo:2019cyg,Caputo:2019xum} the same problem has been solved using effective field theory (EFT) techniques for the description of collective excitations in different phases of matter---see, e.g.,~\cite{Son:2002zn,Nicolis:2013lma,Nicolis:2015sra,Nicolis:2017eqo}. This allows to bypass the complicacies of the microscopic physics of the detector, formulating a low-energy quantum field theory with a given symmetry breaking pattern. 
Amplitudes and rates can be computed with perturbation theory methods.

\begin{figure}[t]
    \centering
    \includegraphics[width=0.75\columnwidth]{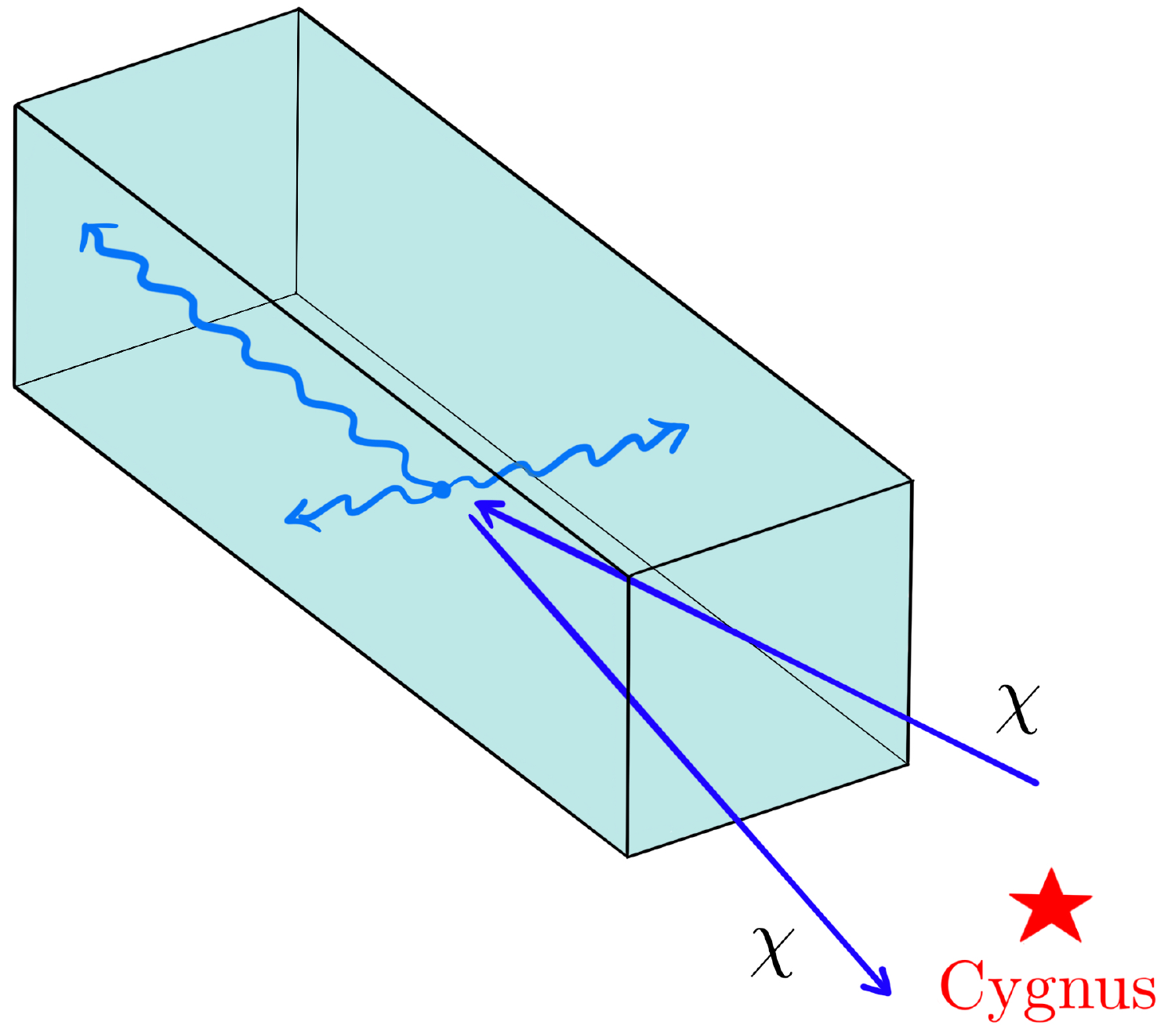}
\caption{Possible ``cygnus-shaped'' event. One phonon is emitted forward, almost in the direction of the incoming dark matter, while the other two are emitted back-to-back and orthogonally to it. We assume the geometry of the detector is such that all three phonons are then redirected upward to be detected via quantum evaporation.} \label{fig:event}
\end{figure}

In this work we use the same approach to investigate a new possible signature of the interaction of dark matter with the \He detector---the emission of three phonons.
This shows the capabilities of the methods used, and paves the way to further discussions on the experimental signatures
in superfluid targets. Indeed, as we will argue, there is a region of mass around roughly $500\keV$ where the three phonons are emitted in the configuration shown in Figure~\ref{fig:event}, which we dub as the ``cygnus'' configuration. {
The details of the experimental setup will tell how to detect these configurations.
}

Although suppressed with respect to other processes~\cite{Schutz:2016tid,Knapen:2016cue,Acanfora:2019con,Caputo:2019cyg,Caputo:2019xum,Baym:2020uos}, this event has the potential to allow for the reconstruction of the dark matter interaction vertex, while providing a good background discrimination via coincidence requirements and directionality. The latter also provides a handle to determine the dark matter mass. In particular, we show that, for a good fraction of the events, the dark matter releases most of its available momentum to the forward phonon, whose direction is then strongly correlated to the direction of the incoming particle.

We present the phonon self-interactions up to quartic order in the field, and at leading order in the small momenta/long wavelength expansion. We  develop an analytic treatment of the four-body phase space to compute the emission rate of three phonons, along with a fully numerical phase space Monte Carlo tool which serves for producing  the results presented in this paper.

\vspace{1em}

\noindent\emph{Conventions:} Throughout this paper we work with a ``mostly plus'' metric and set $\hbar=c=1$.


\section{Effective action} \label{sec:LO}

When the momentum exchanged by the dark matter to the detector is smaller than the inverse of the typical atomic size, the dark matter cannot resolve single atoms but rather it interacts with macroscopic collective excitations. A particularly suitable description of the latter is in terms of EFTs, which are independent of the (often complicated) microscopic details of the condensed matter system and rely solely on symmetry arguments.

In particular, the last decade witnessed the development of relativistic EFTs for different phases of matter, which are based on the observation that condensed matter systems are particular symmetry violating states of an underlying Poincar\'e invariant theory, which is then spontaneously broken (see, e.g.,~\cite{Nicolis:2013lma,Nicolis:2015sra}). The soft collective excitations of the system are nothing but the corresponding Goldstone bosons, whose interactions are strongly constrained by the nonlinearly realized symmetries, and can be organized in a derivative (low-energy) expansion.

From this viewpoint, a superfluid like \He is a system that spontaneously breaks boosts, time translations and an internal $U(1)$ symmetry associated to particle number conservation, but preserves a diagonal combination of the last two (see, e.g.,~\cite{Son:2002zn,Nicolis:2011cs,Nicolis:2015sra,Nicolis:2017eqo}). The most general effective Lagrangian, at leading order in the low-energy expansion, is given by $\mathcal{L}=P\left(X\right)$, where $P$ is the pressure of the superfluid as a function of the chemical potential, $X=\sqrt{-\partial_\mu\psi \partial^\mu\psi}$, and $\psi(x)=\mu t + \sqrt{\mu c_s^2/\bar n}\,\pi(x)$, with $\pi(x)$ being the phonon field. The parameters $c_s$, $\mu$ and $\nbar$ are, respectively, the equilibrium sound speed, relativistic chemical potential and number density. For more details we refer the reader to~\cite{Acanfora:2019con} and the references therein.

The action up to quartic order in the field reads
\begin{align} \label{eq:SLOph}
    \begin{split}
        S_\text{ph}&=\int dtd^3x\bigg[ \frac{1}{2}\dot\pi^2 - \frac{c_s^2}{2}\left( \del \pi \right)^2 \\
        & \quad + \frac{g_1}{2} \dot\pi\left( \del\pi \right)^2 + \frac{g_2}{6}\dot\pi^3 \\
        & \quad + \frac{\lambda_1}{8} \left( \del\pi\right)^4 + \frac{\lambda_2}{4}\dot\pi^2\left(\del\pi\right)^2 + \frac{\lambda_3}{24}\dot\pi^4 \bigg]\,.
    \end{split}
\end{align}
The effective couplings in the nonrelativistic limit\footnote{We remark that, although the dark matter itself is characterized by small speeds, here we mean the nonrelativistic limit of a superfluid like \He, i.e., the instance where $c_s\ll 1$ and $\mu\simeq \mHe$.} are related to the thermodynamic quantities of the superfluid by the following expressions~\cite{Acanfora:2019con}:
\begin{align}
    g_1 &\simeq -\sqrt{\frac{\mHe \cs^2}{\nbar}}\frac{1}{\mHe}\,, \quad g_2 \simeq \left(\frac{\mHe \cs^2}{\nbar}\right)^\frac{3}{2} \nbar''\,, \\
    \lambda_1 &\simeq \frac{\cs^2}{\mHe \nbar}\,, \quad  \lambda_2 \simeq - \frac{\mHe \cs^4}{\nbar^2} \nbar''\,, \quad \lambda_3 \simeq \frac{\mHe^2 \cs^4}{\nbar^2} \nbar'''\,, \notag
\end{align}
where the primes denote derivatives with respect to the chemical potential.
In Table~\ref{tab:params} we report the values of the above quantities for \He, as obtained from the equation of state reported in~\cite{caupin2008static}.\footnote{We notice that the parameter $b$ in the Appendix of~\cite{caupin2008static} has a typo in its units, which should be $\text{MPa}^{1/2}\,\text{m}^{9/2}\,\text{kg}^{-3/2}$, as also confirmed by the data reported in Figure~6 of~\cite{campbell2015dynamic}.}

\bgroup
\def\arraystretch{1.4}
\begin{table*}[t]
    \begin{tabular}{c|c|c|c|c}
        \hline\hline
        $\mHe$ & $\cs$ & $\nbar$ & $\nbar''$ & $\nbar'''$ \\
        \hline
        \;\;$3.75 \times 10^{6} \keV$\;\; & \;\;$8.2 \times 10^{-7}$\;\; & \;\;$0.17 \keV^3$\;\; & \;\;$-1.16 \times 10^{11} \keV$\;\; & \;\;$5.47 \times 10^{17}$\;\; \\
        \hline\hline
    \end{tabular}
    \caption{Thermodynamic quantities for \He at atmospheric pressure, as obtained from the equation of state reported in~\cite{caupin2008static}. Derivatives with respect to the chemical potential can be related to those with respect to the pressure by the identity $dP = \nbar d\mu$.} \label{tab:params}
\end{table*}
\egroup

In the class of models we are considering, the effective coupling with the dark matter in the nonrelativistic limit occurs via the number density operator,\footnote{This is true for the most common models~\cite{Knapen:2017xzo}, where the coupling to the Standard Model happens either via a current-current interaction or via the trace of the stress energy tensor, which both reduce to the number density in the nonrelativistic limit.} {$\mathcal{L}_\text{int} = G_\chi \mchi \left|\chi\right|^2 n(X)$. The latter is easily found as the temporal component of the Noether current associated with the superfluid $U(1)$ symmetry, $\psi\to\psi+a$, with $a$ being constant}. This leads to the following interaction term~\cite{Acanfora:2019con,Caputo:2019cyg,Caputo:2019xum}
\begin{align} \label{eq:SLOint}
    \begin{split}
        S_\text{int} &= \int dt d^3x \,G_\chi \mchi \left|\chi\right|^2 \bigg[ -\alpha \dot\pi - \frac{\beta_1}{2} \left( \del\pi\right)^2 \\ 
        & \quad - \frac{\beta_2}{2}\dot\pi^2 + \frac{\gamma_1}{2} \dot\pi\left(\del\pi\right)^2 + \frac{\gamma_2}{6} \dot\pi^3 \bigg]\,,
    \end{split}
\end{align}
where $\mchi$ is the dark matter mass and $G_\chi$ an effective coupling of the dark sector with the dimension $(\text{mass})^{-2}$, which can eventually be  related to the dark matter--nucleon cross section by the relation $\sigma_n = G_\chi^2 \mu_{\chi n}^2/(4\pi)$, with $\mu_{\chi n}$ being their reduced mass. We consider the case of a complex scalar dark matter given that, for small dark matter velocity, the final results are spin independent. The effective couplings can again be related to the quantities in Table~\ref{tab:params}:
\begin{align}
    \begin{split}
        \alpha \simeq -\sqrt{\frac{\nbar}{\mHe\cs^2}}\,, \quad \beta_1 \simeq& \frac{1}{\mHe}\,, \quad \beta_2 \simeq -\frac{\mHe \cs^2}{\nbar} \nbar''\,, \\
        \gamma_1 \simeq -\left(\frac{\mHe \cs^2}{\nbar}\right)^\frac{3}{2} \!\!\frac{\nbar''}{\mHe}\,,& \quad \, \gamma_2 \simeq \left(\frac{\mHe \cs^2}{\nbar}\right)^\frac{3}{2} \nbar'''\,.
    \end{split}
\end{align}

The actions \eqref{eq:SLOph} and \eqref{eq:SLOint} produce the following Feynman rules for the self-interaction of phonons of energy and momentum $(\omega,\bm q)$,
\begin{align*}
    \includegraphics[width=0.12\textwidth,valign=c]{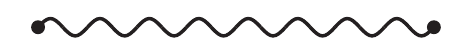} &= \frac{i}{\omega^2 - \cs^2 \, {\bm{q}}^2 + i\varepsilon}\,, \\
    \includegraphics[width=0.12\textwidth,valign=c]{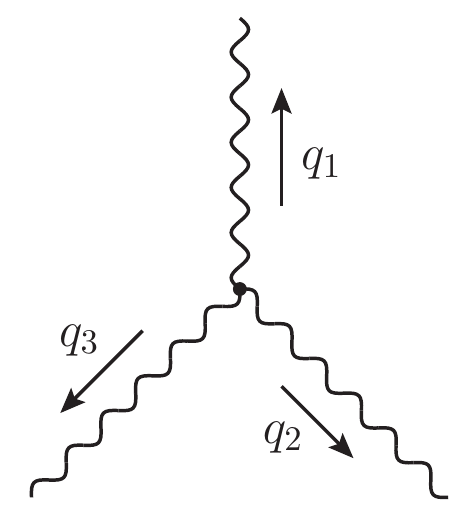} &= g_1\big( \omega_1 \, \sprod{2}{3} + \omega_2 \, \sprod{1}{3} + \omega_3 \, \sprod{1}{2} \big) \\[-2.25em]
    &\quad + g_2 \, \omega_1\,\omega_2\,\omega_3 \,, \\[1em]
    \includegraphics[width=0.12\textwidth,valign=c]{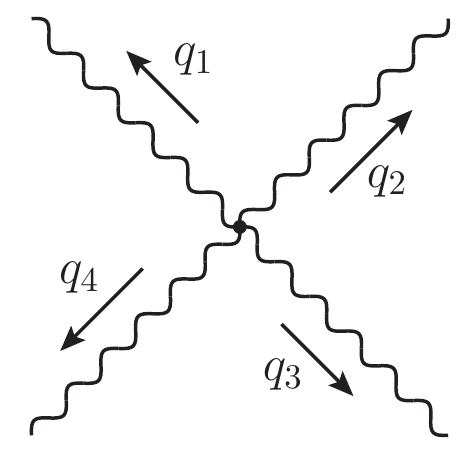} &= i\lambda_1 \big( \sprod{1}{2}\,\sprod{3}{4} + \sprod{1}{3}\,\sprod{2}{4} \\[-2.25em]
    &\qquad + \sprod{1}{4}\,\sprod{2}{3} \big) \\
    &\quad + i\lambda_2 \big( \omega_1\,\omega_2\,\sprod{3}{4} + \omega_1\,\omega_3\,\sprod{2}{4} \\
    &\quad\quad\;\;\; + \omega_1\,\omega_4\,\sprod{2}{3} + \omega_2\,\omega_3\,\sprod{1}{4} \\
    &\quad\quad\;\;\; + \omega_2\,\omega_4\,\sprod{1}{3} + \omega_3\,\omega_4\,\sprod{1}{2} \big) \\
    &\quad + i\lambda_3\,\omega_1\,\omega_2\,\omega_3\,\omega_4\,,
    \end{align*}
and for their interaction with dark matter,
\begin{align*}
    \includegraphics[width=0.14\textwidth,valign=c]{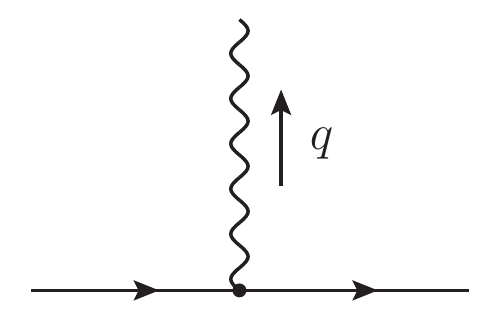} &= G_\chi \mchi \alpha \,\omega\,, \\
    \includegraphics[width=0.14\textwidth,valign=c]{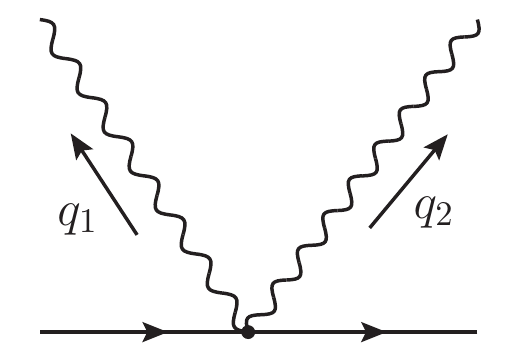} &= i G_\chi \mchi \big( \beta_1 \sprod{1}{2} + \beta_2 \omega_1\,\omega_2 \big)\,, \\
    \includegraphics[width=0.14\textwidth,valign=c]{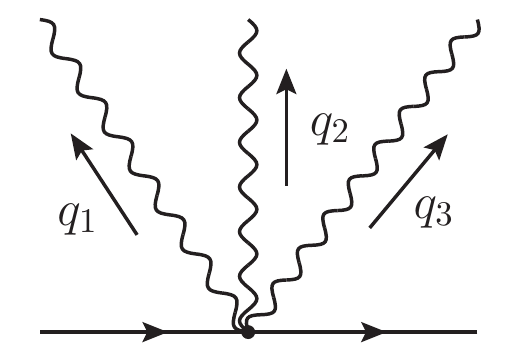} &= G_\chi \mchi \big[ \gamma_1 \big( \omega_1\,\sprod{2}{3} + \omega_2\,\sprod{1}{3} \\[-1.8em]
    & \qquad \qquad \;\;\, + \omega_3\,\sprod{1}{2} \big) + \gamma_2 \, \omega_1\,\omega_2\,\omega_3 \big]\,.
\end{align*}

Let us stress that, the EFT being a low-energy theory, it is valid up to a certain strong coupling scale, $M_\text{UV}$. At momenta higher than that, the phonon becomes strongly coupled, and the derivative expansion breaks down. For \He, $M_\text{UV}$ can be deduced, for example, from the value of the momentum for which the dispersion relation deviates from linear, $\omega = c_s |\bm{q}|$, by order one corrections, or for which the phonon width is of the same order as its frequency. To ensure the perturbativity of our treatment we then limit all momenta to be smaller than a cutoff, $\Lambda=1\keV$. At momenta equal to $\Lambda$, indeed, the corrections to the derivative expansion are still moderate, as proved by the fact that the deviation from the linear dispersion relation is $\simeq30\%$~\cite{maris1977phonon}, and that the phonon width compared to its frequency is $\Gamma/\omega\simeq3\%$. The effects of nonlinearities in the dispersion relation for the problem at hand have been  discussed via standard techniques in~\cite{Baym:2020uos}.


\section{Phase space for three-phonon emission} \label{sec:3phonon}

It is now possible to compute the rate of emission of three phonons by the passing dark matter. We will assume that all phonons are separately detected via quantum evaporation~\cite{Bandler:1992zz,Maris:2017xvi,Hertel:2018aal,Osterman:2020xkb}, making them distinguishable from each other. This is possible if each of them has energy larger than the binding energy of a helium atom to the surface of the superfluid, i.e., if $\omega_i\geq0.62\meV$~\cite{Hertel:2018aal}. Moreover, when $\omega_i > 0.68\meV$, phonons are stable against decay into two other phonons~\cite{maris1977phonon,Hertel:2018aal}. We impose that all final state phonons satisfy this condition.

\begin{figure}[t]
    \centering
    \includegraphics[width=0.35\columnwidth]{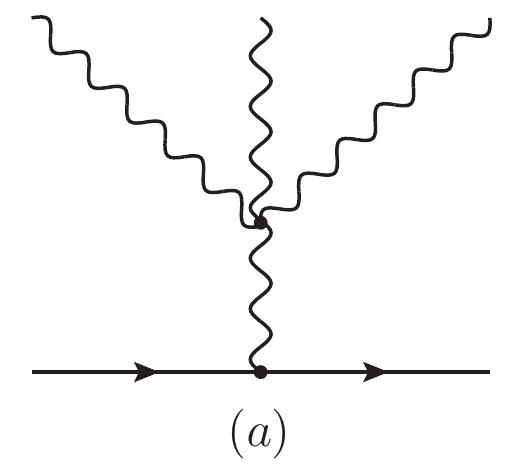}
    \hspace{1em}
    \includegraphics[width=0.35\columnwidth]{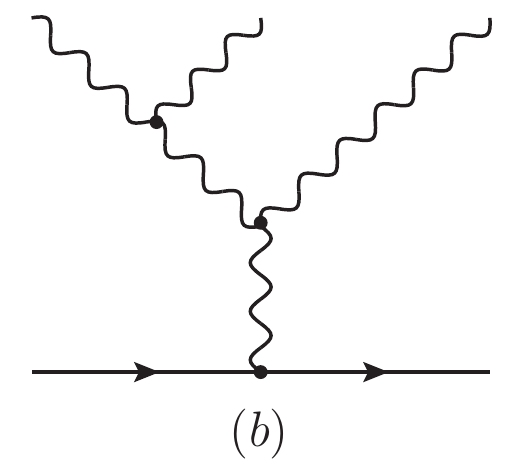}
    \hspace{1em}
    \includegraphics[width=0.35\columnwidth]{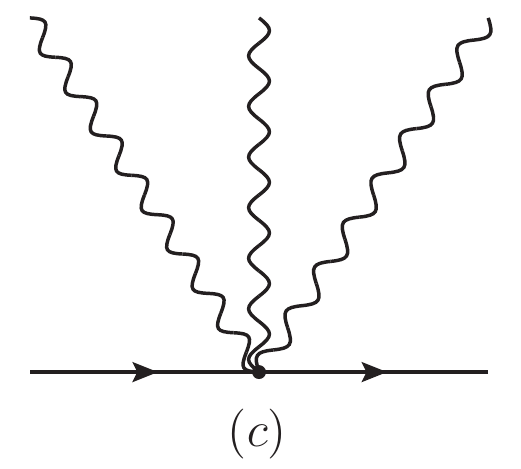}
    \hspace{1em}
    \includegraphics[width=0.35\columnwidth]{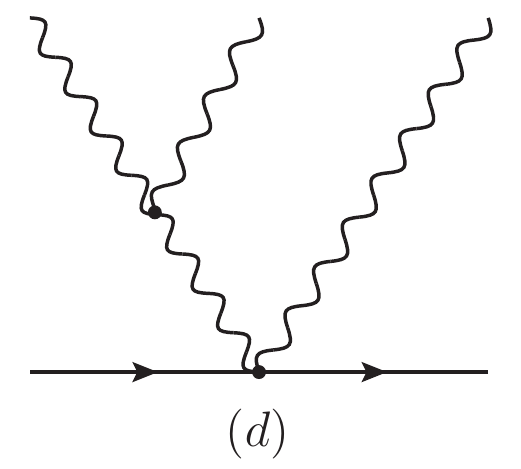}
    \caption{Diagrams contributing to the emission of three phonons. Permutations of the external momenta for the diagrams $(b)$ and $(d)$ are understood.} \label{fig:3phonon_diags}
\end{figure}

The amplitude for the process under consideration is given by the diagrams in Figure~\ref{fig:3phonon_diags}, with appropriate permutations of the external momenta.
We recall that, for small total exchanged momentum $\bm q$, the overall amplitude is strongly suppressed, $\mathcal{M}=O(q^2)$~\cite{Knapen:2016cue,Caputo:2019cyg}, as a consequence of the conservation of particle number associated to the superfluid~\cite{Caputo:2019ywq} (see also~\cite{Baym:2020uos}). In this particular case, this happens via a pairwise cancellation between the diagrams $\mathcal{M}_a$ and $\mathcal{M}_c$, and $\mathcal{M}_b$ and $\mathcal{M}_d$. {This can be understood in terms of integrating out the highly off-shell intermediate phonon, which amounts to shrinking its propagator to a pointlike local interaction~\cite{Caputo:2019cyg}. The expression for the matrix element is admittedly cumbersome, but nonetheless  manageable analytically. We report it in Appendix~\ref{app:M}, together with some further discussion.}

We now present a semi-analytical treatment of the four-body phase space associated to the three-phonon emission rate, which we compare to a full Monte Carlo calculation.

When the light dark matter particle hits the superfluid target in a specific point, the helium volume
reacts as a whole, and one or more phonons are produced in the neighbourhood of the interaction point. Within the EFT, this is described in terms of an elementary pointlike process in which the dark matter--helium interaction generates a given number of phonons.   

For simplicity, let us first consider the two-phonon case. In its passage, the light dark matter particle exchanges a {\it spacelike momentum}\footnote{In a Lorentz-breaking medium spacelike, timelike and lightlike are no longer good labels. However, it is still useful to phrase the process in these terms to help intuition, and to implement the phase space with Monte Carlo techniques.} and the superfluid \He responds with the excitation of  two gapless particles; including the final state dark matter, we have a $1\to 3$ process. 
In Minkowski metric, on-shell phonons are spacelike:  if we compose a phonon 4-momentum in the form $q_\mu=(\omega,\bm q)$,  this follows from the dispersion law $\omega =c_s |\bm{q}|$, with $c_s\sim 10^{-6}$. 

To compute the phase space, we factorize the $1\to 3$ process (dark matter $\to$ dark matter $+$ two phonons) into two parts
\begin{align*}
    p \to p_1 + q_1 + q_2 = \big( p \to p_1 + q \big) \times \big( q \to q_1 + q_2 \big)\,.
\end{align*}
We will use $p$ for the norm of the momentum $\bm p$ of the incoming dark matter particle, $p_1$ for that of the outgoing one and $q_i$ for the final state phonons.
The 4-momentum $q$ is fictitious.
Formally, we treat the first factor as a two-body decay into the final state dark matter particle and a ``tachyon", as described by the phase space integral
\begin{align}
    I_2=\frac{1}{16\pi^2}\int \frac{d^3p_1}{\mchi}\frac{d^3q}{\sqrt{q^2+s}}\delta^4(p-p_1-q)\,,
\end{align}
with 
\begin{align} \label{basicineq}
    q^2 \geq -s\,, \qquad \text{and} \qquad s = \omega^2 - q^2\,,
\end{align} 
where $\omega = \frac{p^2}{2\mchi}-\frac{p_1^2}{2\mchi}$ is the energy released to the system by the dark matter particle.
From now on, we work at leading order in the nonrelativistic limit, $c_s\ll1$.
The $I_2$ integral can be written as
\begin{align}
    \begin{split}
        I_2 &= \frac{1}{16\pi^2}\int \frac{d^3p_1}{\mchi\sqrt{(\bm p-\bm p_1)^2+s}} \\
        &\quad\times \delta\left(\omega -\sqrt{p^2+p_1^2-2 p p_1 \cos\theta+s}\right)\,.
    \end{split}
\end{align}
Solving the $\delta$-function for $\cos\theta$, and requiring it to have support on the integration region, we get
a condition on the possible values of $s$, i.e., $s_{\rm min} \leq s \leq s_{\rm max}$ with
\begin{align}
    s_{\rm min} = \omega^2 - {(p+p_1)}^2\,, \quad s_{\rm max} = \omega^2 - {(p-p_1)}^2\,.
\end{align}

The three-body phase space volume $I_3$ must include the phase space for the decay of the ``tachyon'' into two phonons, which we call $J_2$. Altogether we get
\begin{align}
    I_3=\frac{1}{8\pi \mchi p} \int_0^{p} p_1\,dp_1 \times \frac{1}{2\pi}\int_{s_{\rm min}}^{s_{\rm max}} ds\times J_2 \,.
\label{iphs}
\end{align}
Now compute the factor $J_2$:
\begin{align} \label{jformula}
    J_2 &= \frac{1}{16\pi^2c_s^2} \int \frac{d^3q_1}{q_1}\frac{d^3q_2}{q_2}\delta^4\left( q - q_1 - q_2 \right) \\
    &= \frac{1}{8\pi c_s^2} \int \frac{dq_1 d\cos\eta\, q_1}{|\bm{q}-\bm{q}_1|}\delta\big( \omega - c_s q_1 - c_s |\bm q - \bm q_1| \big)\,, \notag
\end{align}
where $\eta$ is the relative angle between $\bm q$ and $\bm q_1$.
Integrating over it, we find
\begin{align}
    J_2=\frac{1}{8\pi c_s^3 q}\int_{q_1^{\rm min}}^{q_1^{\rm max}}dq_1\,.
\end{align}
Imposing $-1\leq \cos\eta\leq 1$, we get
\begin{align}
    q_1^{\rm min}=\frac{\sqrt{q^2+s}}{2 c_s}-\frac{q}{2}\,, \qquad
    q_1^{\rm max}=\frac{\sqrt{q^2+s}}{2 c_s}+\frac{q}{2}\,.
\end{align}
Plugging everything into Eq.~\eqref{iphs} gives the phase space volume for the three-body decay:
\begin{align} \label{eq:I3}
    I_3 = \frac{p^3}{96\pi^3 c_s^3 \mchi }\,.
\end{align}

\begin{figure}[t]
    \centering
    \includegraphics[width=0.75\columnwidth]{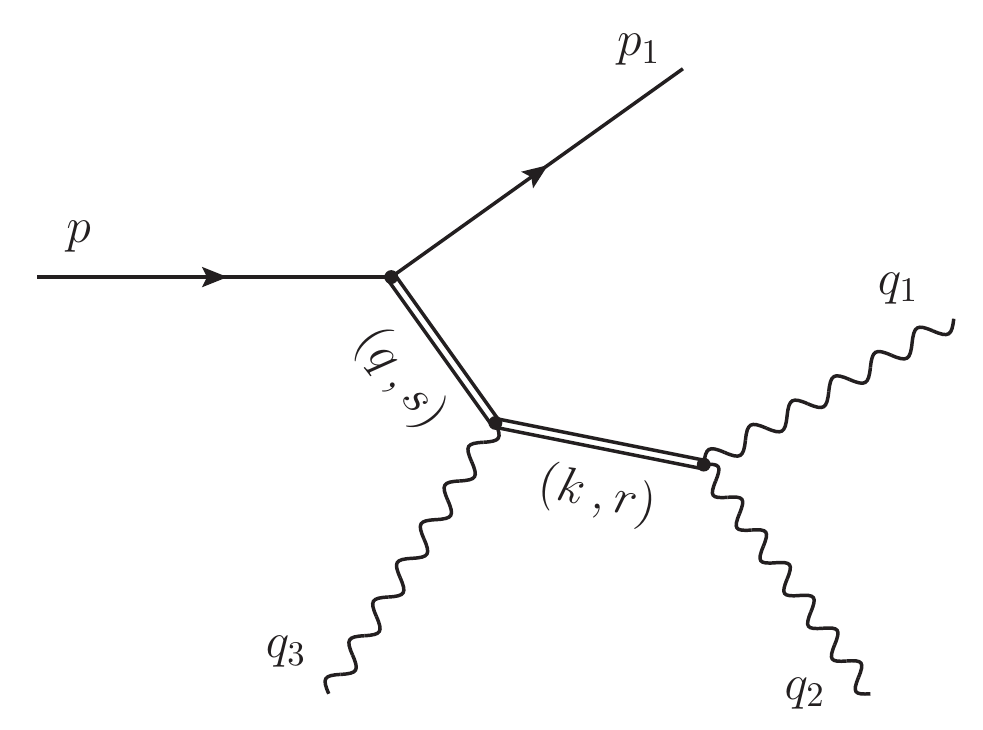}
    \caption{Kinematics of three-phonon emission. The $1\to4$ process is factorized in three $1\to2$ processes by introducing two fictitious 4-momenta.} \label{fig:3phonon_kin}
\end{figure}

The calculation of the four-body phase space follows the same logic, but is admittedly more involved. In particular, the decay is now factorized in three two-body decays, with the introduction of two fictitious 4-momenta, as shown in Figure~\ref{fig:3phonon_kin}.
We now have two integration variables, $s$ and $r$, and the result for the four-body phase space factor, $I_4$, is 
\begin{align}
    \begin{split}
        I_4&= \frac{1}{8\pi \mchi p} \int_0^{p}\,p_1dp_1 \times \frac{1}{2\pi}\int_{s_{\rm min}}^{s_{\rm max}} ds \\
        &\quad\times \frac{1}{8\pi c_s q} \int_0^{q_3^{\rm max}}\, dq_3 \times \frac{1}{2\pi} \int_{r_{\rm min}}^{r_{\rm max}}dr  \\
        &\quad\times \frac{1}{16\pi^2c_s^3k} \int_{q_1^{\rm min}}^{q_1^{\rm max}}dq_1 \int_0^{2\pi} d\phi_{1}\,,
    \end{split}
\end{align}
where $\phi_1$ is the azimuthal angle between $\bm{q}_1$ and $\bm k$.
Imposing positivity of all energies, and for the $\delta$-functions to have support within the integration domain, one finds the following additional extrema:
\begin{subequations}
    \begin{align}
        r_{\rm min} &= s - q_3^2 - 2 c_s q_3 \omega - 2 q q_3\,,  \\
        r_{\rm max} &= s - q_3^2 - 2 c_s q_3 \omega + 2 q q_3\,, \\
        q_3^{\rm max} &= \frac{\omega}{2c_s}\,.
    \end{align}
\end{subequations}
Putting everything together, the expression for the four-body phase space is
\begin{align} \label{eq:I4}
I_4 = \frac{p^7}{53760\, \mchi^3 \pi^5 c_s^6}\,.
\end{align}

One last comment is in order. The matrix element for the three-phonon emission is a function of the magnitudes of the phonon momenta, $q_i$, and of their relative polar angles, $\theta_{ij}$. After imposing momentum conservation to, say, eliminate $\bm{q}_2$, one is still left with a dependence on $\theta_{13}$, which does not appear among the integration variables. The latter can, however, be related to the relative angles between $\bm q_1$ and $\bm k$, say $(\theta_1,\phi_1)$, and between $\bm q_3$ and $\bm k$, say $(\theta_3,\phi_3)$. To do that, start from a reference frame where $\bm{q}_1$ is along the $z$-axis. The other two vectors can be written as $\bm q_3=q_3(\sin\theta_{13}\cos\phi_{13},\sin\theta_{13}\sin\phi_{13},\cos\theta_{13})$ and $\bm k = k(\sin\theta_1\cos\phi_1,\sin\theta_1\sin\phi_1,\cos\theta_1)$. We now perform a rotation to a frame where $\bm k$ is along the $z$-axis. The corresponding rotation matrix is given by $R = e^{\bm{L}\cdot\hat{\bm{n}}\,\theta_1} = \mathbb{1} + \sin\theta_1 \bm{L}\cdot\hat{\bm{n}} +  (1-\cos\theta_1)\left(\bm{L}\cdot\hat{\bm{n}} \right)^2$, where $\bm{L}$ are the real generators of three-dimensional rotations, and $\hat{\bm{n}} = \bm k \times \bm q_1 / \left| \bm k \times \bm q_1 \right|$ is the normal vector perpendicular to both $\bm k$ and $\bm q_1$. In this frame, the momentum $\bm q_3$ is given by $R\cdot \bm q_3 = q_3 (\sin\theta_3\cos\phi_3,\sin\theta_3\sin\phi_3,\cos\theta_3)$. Rotating this last vector back to the original frame, one finds the relative angle between $\bm q_1$ and $\bm q_3$ as a function of the angles between $\bm q_1$ and $\bm k$, and $\bm q_3$ and $\bm k$, i.e.,
\begin{align} \label{eq:costheta13}
    \cos\theta_{13} = \cos \theta_3 \cos\theta_1 - \cos\big(\phi_3-\phi_1\big)\sin\theta_3\sin\theta_1\,.
\end{align}
All the remaining angles are either trivial, an integration variable, or can be eliminated through a $\delta$-function.

The analytic formulae for phase space volumes in  Eqs.~\eqref{eq:I3} and \eqref{eq:I4}, {valid at leading order in $c_s\ll1$}, have been compared to the fully numerical {exact} computation of the same two quantities, finding perfect agreement.

The numerical calculation proceeds through the random generation of final state momenta and their selection with a sequence of controls cutting the phase space according to the kinematical conditions. 
The matrix element, as determined by the EFT, will weight the phase space cells. The differential distributions presented here are generated numerically.


\section{Differential rates, vertex reconstruction and directionality}

\begin{figure*}[t]
    \centering
    \includegraphics[width=0.495\textwidth]{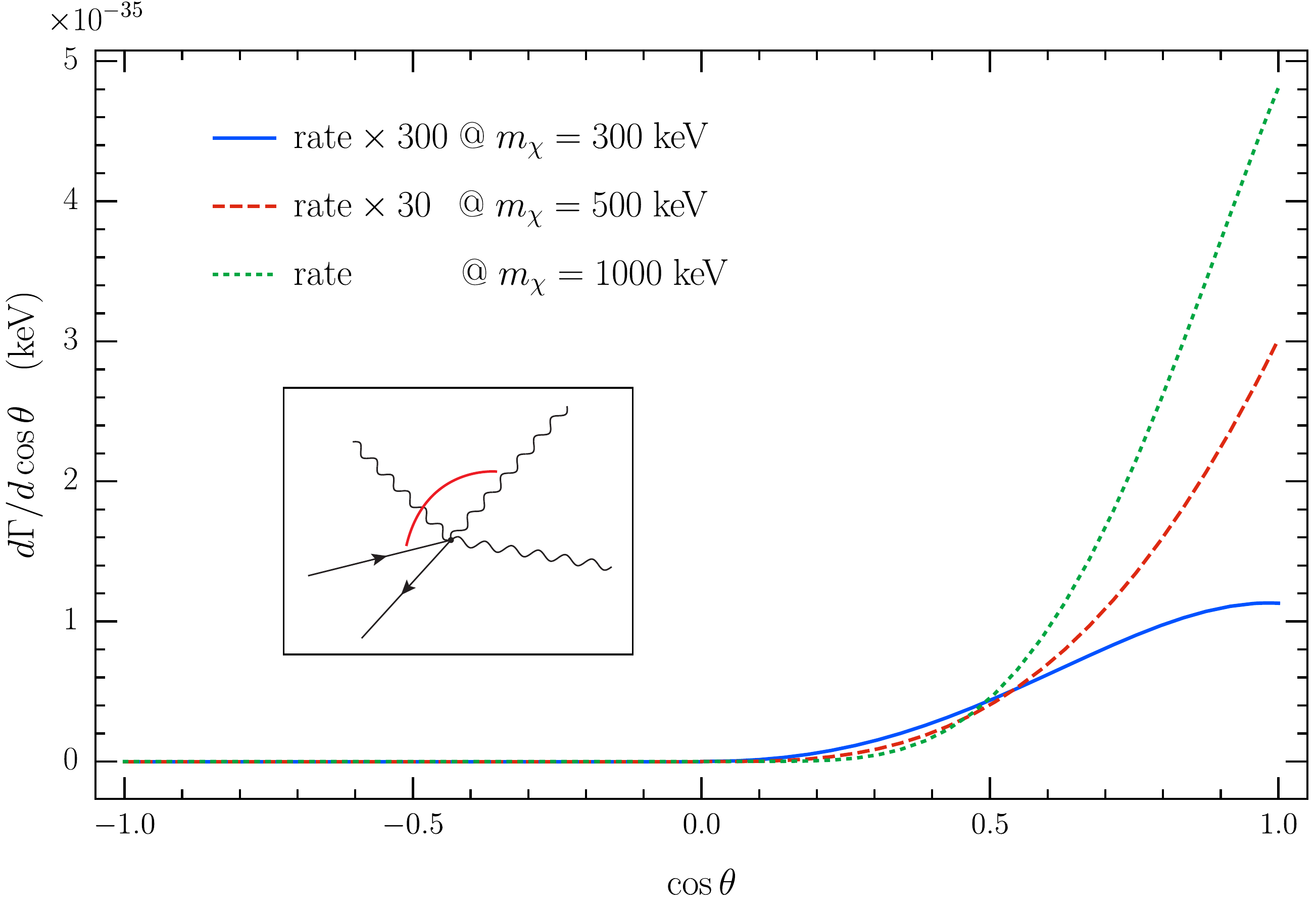}
    \includegraphics[width=0.495\textwidth]{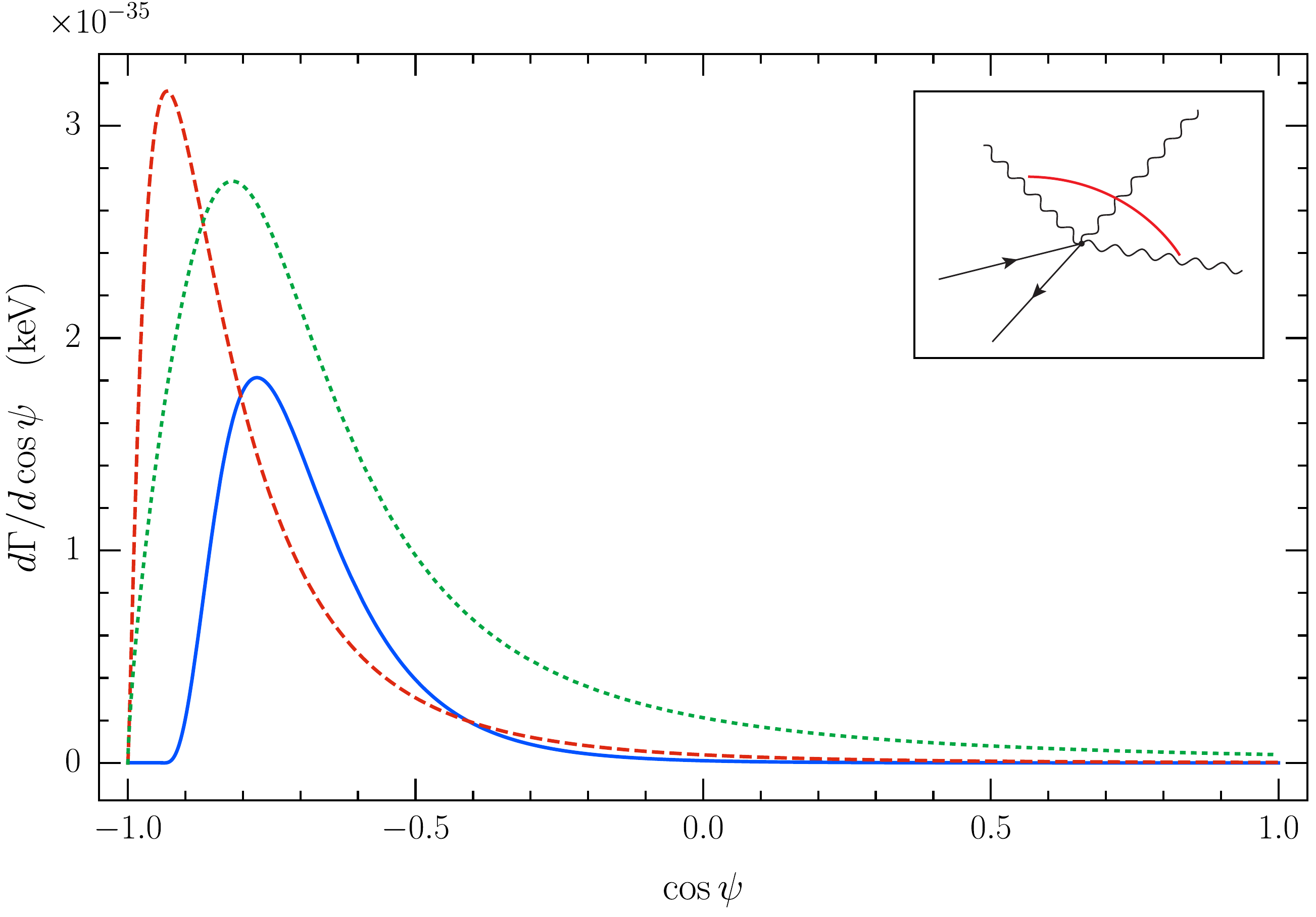}
    \includegraphics[width=0.495\textwidth]{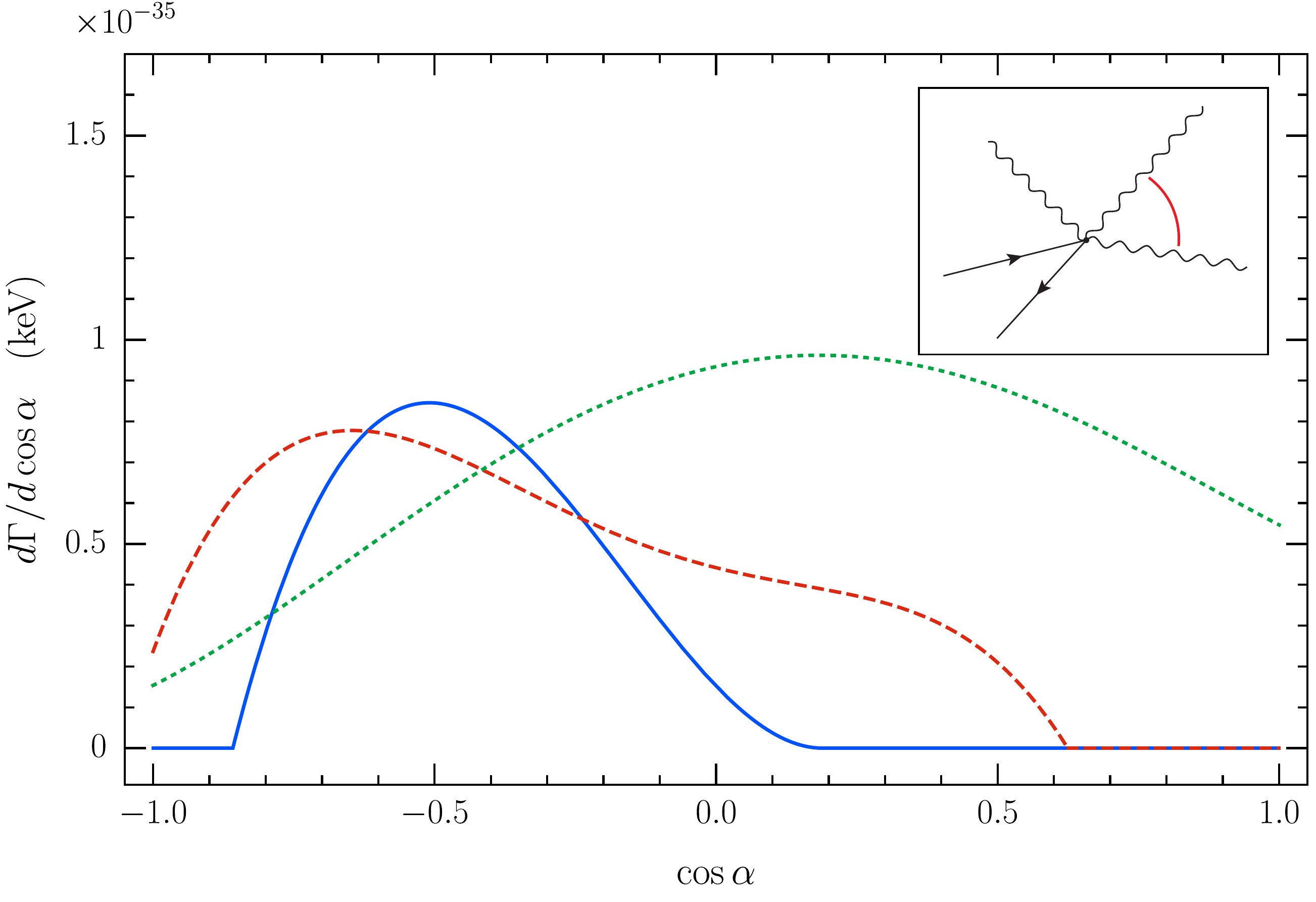}
    \includegraphics[width=0.495\textwidth]{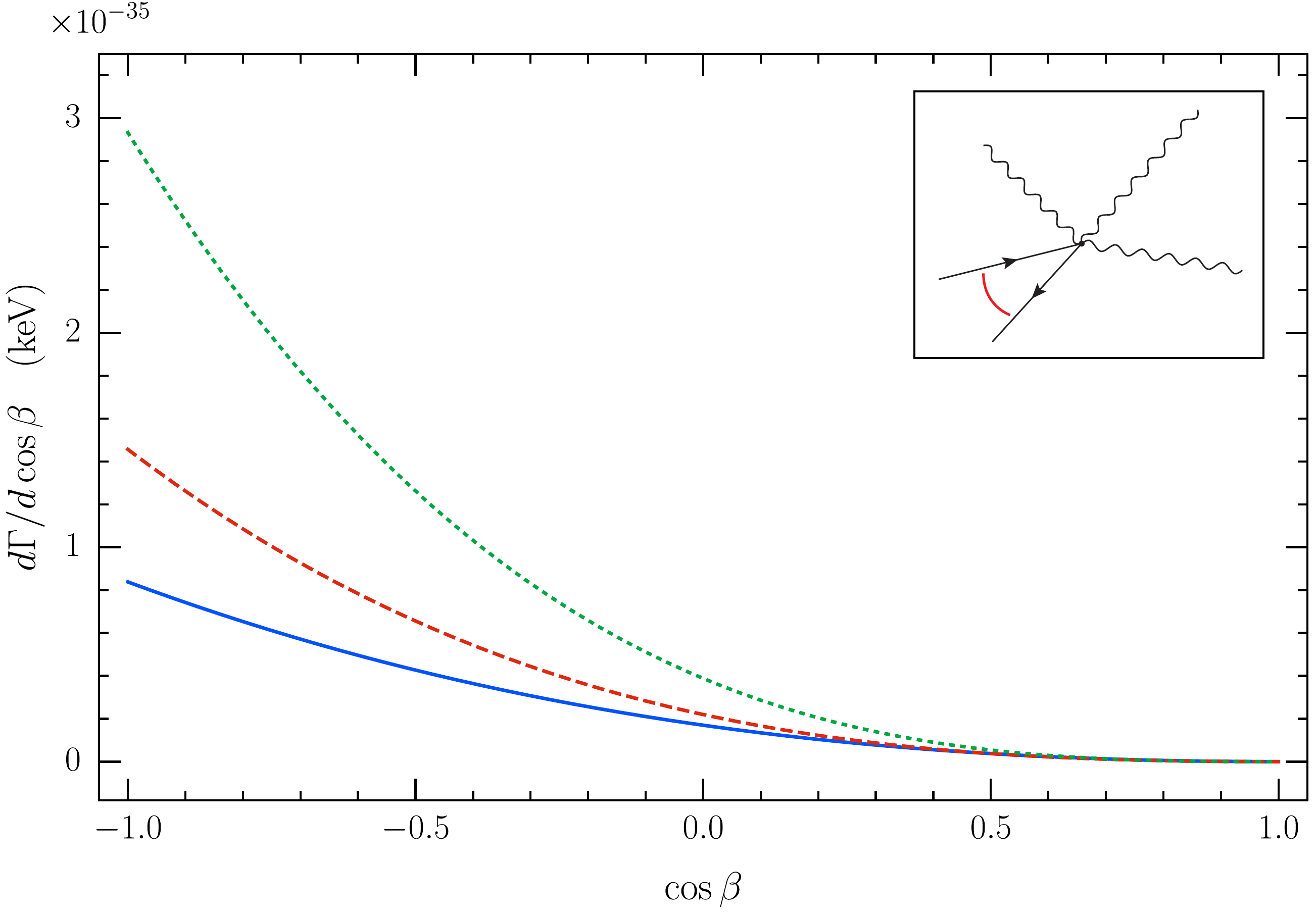}
\caption{Differential distributions for the cosine of the angle between the most forward phonon and the direction of the incoming dark matter (upper left), the angle between the other two phonons (upper right), the angle between the most forward phonon and the hardest of the remaining two (bottom left) and the angle between the incoming and outgoing dark matter (bottom right). The corresponding angles are pictorially represented in the insets. Note that, in the top right panel, while going from $\mchi=300\keV$ to $500\keV$, the typical angle gets closer to $\psi=\pi$ and further increasing the mass to $1000\keV$ brings it back to smaller values. This is due to an interplay between the high momentum released by the dark matter, and the presence of the cutoff $\Lambda$ for the forward phonon. The rates have been obtained for a nominal value of the dark matter--nucleon cross section $\sigma_n = 10^{-42}$~cm$^2$. We recall that, in a realistic situation, the dark matter incoming direction follows a Maxwell-Boltzmann distribution. The curves for $\mchi=300\keV$ and $500\keV$ have been rescaled to make their shape visible, as indicated in the legend.} \label{fig:dist}
\end{figure*}

The relevance of an event where three phonons are emitted lies in its potential for background rejection, vertex reconstruction, and directionality. Consider the ``cygnus-shaped'' event, like the one schematically represented in Figure~\ref{fig:event}.
The corresponding rate will clearly be suppressed with respect to that of emission of a lower number of phonons. Nonetheless, {one can envision a detector where} all three phonons can be observed and where their direction can be approximately determined (see~\cite{enss1994quantum,Maris:2017xvi}). {This might be achievable, for example, with a geometry as the one depicted in Figure~\ref{fig:event}, with the vertical sides tilted in order to redirect the two back-to-back phonons upward, allowing them to trigger quantum evaporation. The latter, as found in~\cite{enss1994quantum,Maris:2017xvi}, only happens if the phonon has a relative angle with respect to the direction normal to the surface of the superfluid smaller than a certain critical value (roughly 25$^\circ$). This can allow for a partial reconstruction of the phonon initial direction. Given this,} an event of this sort contains important information, which cannot be obtained from other processes.
First, the presence of two almost back-to-back phonons could be employed for efficient background rejection through coincidence requirements~\cite{Hertel:2018aal}, {also in combination with the third forward phonon}.
Moreover, knowing the direction of the outgoing phonons, it is possible to reconstruct the interaction vertex. {Used in combination with timing information,\footnote{{With a time resolution of $\Delta t\sim0.1$~ms, as the one envisioned in~\cite{Hertel:2018aal}, one can in principle discriminate distances of the order of $\Delta x\sim c_s\Delta t\sim 2$~cm.}} this can be important to discriminate between a multiphonon emission due to a single scattering with the target (as expected for a dark matter particle) and one due to multiple scatterings (as it can happen, for example, from a background neutron).} As we show below, the direction of the forward outgoing phonon is {also} strongly correlated with the direction of the incoming dark matter, hence providing important directional information, also key to background discrimination. {An event of this sort contains a good deal of information---see Figure~\ref{fig:dist}---which could be further used to characterize the dark matter event.}

To determine whether the configuration shown in Figure~\ref{fig:event} is allowed, we compute different angular distributions---see Figure~\ref{fig:dist}. We consider a typical dark matter velocity given by $v_\chi = 220$~km/s~\cite{lewin1996review}.
Before commenting on them, let us notice that our tree-level amplitudes feature collinear divergences, when two of the final state phonons are emitted at a small relative angle. This is a standard property of amplitudes involving gapless states with a linear dispersion relation. For the phonons of energy between $0.68\meV$ and $c_s\Lambda=0.82\meV$ considered here, the divergence is never hit. Moreover, since we are interested in an exclusive process, where all phonons are well separated in angle, here it suffices to regularize the possible unphysical enhancements including the finite phonon width in the propagator, along the lines of~\cite{Baym:2020uos}---see Appendix~\ref{app:M}.

As one can see from Figure~\ref{fig:dist}, most of the events are such that the forward phonon is indeed close in direction to the incoming dark matter, while the remaining two phonons are almost back-to-back. 
Moreover a non-negligible fraction of the latter appears perpendicular to the forward phonon, i.e., in the configuration of interest to us.
Finally, the differential distribution shows that, in many events, the outgoing direction of the dark matter is almost opposite to the incoming one. The discrepancy between the two can be further reduced by aligning the detector with the direction of the Cygnus constellation, taking advantage of the velocity distribution of the dark matter (see, e.g.,~\cite{Ling:2009eh}). This, together with the presence of the forward phonon, ensure both the directionality of the signal (crucial for background discrimination) and the possibility of reconstructing the dark matter momentum (and hence its mass) from the momentum of the forward phonon.

As we can also deduce from the distributions, the ``cygnus'' event is relevant for masses that are not much lighter than a few hundred keV. Below that, the number of back-to-back phonons emitted perpendicularly to the forward one drops (bottom left panel of Figure~\ref{fig:dist}), making the event unlikely.

We note that the above features cannot be obtained from processes where the dark matter emits one or two phonons. In the former case, while directional information might be available through \v{C}hrenkov emission~\cite{Acanfora:2019con}, this strongly varies with the dark matter mass, it is not possible to reconstruct the interaction vertex, and the process is only allowed for masses strictly heavier than $500\keV$. In the latter case, instead, 
there is a large degeneracy on the interaction point, which makes the reconstruction of the direction of the incoming dark matter, {as well as the interaction vertex}, very challenging. 

\begin{figure}
    \centering
    \includegraphics[width=\columnwidth]{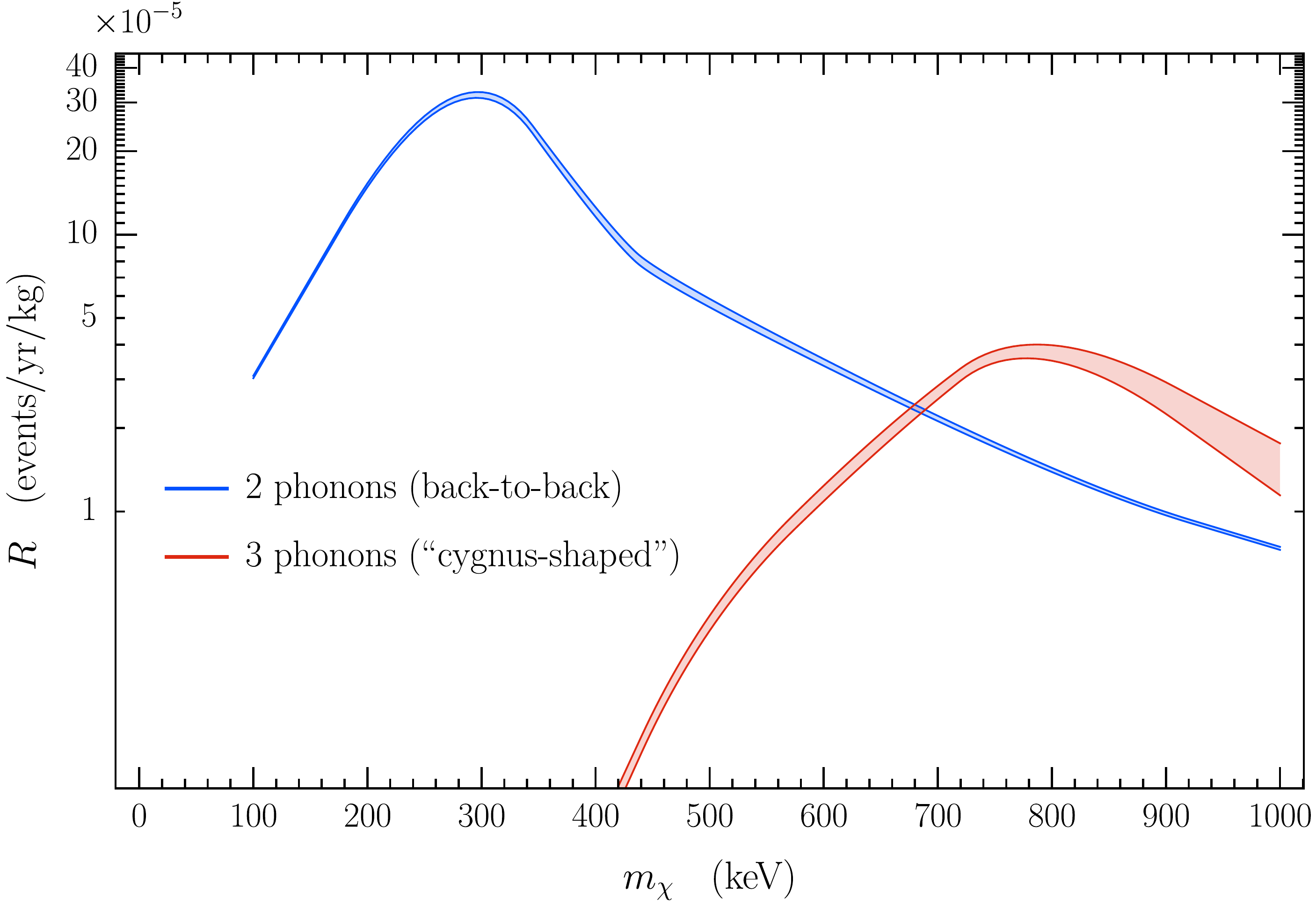}
    \caption{{Number of events per unit time and target mass for two phonons almost back-to-back and three phonons in the ``cygnus-shaped'' configuration. The angular tolerance is taken to be~$25^\circ$~\cite{enss1994quantum,Maris:2017xvi} and the dark matter--nucleon cross section is fixed to the nominal value of $\sigma_n = 10^{-42}$ cm$^2$. The shaded band corresponds to the statistical uncertainty of the Monte Carlo integration. Interestingly, due to the paucity of back-to-back phonons at masses $\mchi\gtrsim 500$~keV, the three-phonon event signature becomes relevant, or even dominant.}}
    \label{fig:comparison}
\end{figure}

To have a rough idea of how many events to expect for a process like this, we estimate the total rate allowing for the three phonons to deviate from the exact ``cygnus-shaped'' configuration by $25^\circ$~\cite{enss1994quantum,Maris:2017xvi} for the relative angle between the dark matter and the forward phonon, between the two back-to-back phonons and between the latter and the forward phonon. {As a comparison we also compute the rate for the emission of two back-to-back phonons, allowing for the same spread in relative angle (essentially the same event but without the additional forward phonon). We fix the dark matter--nucleon cross section to the nominal value of $\sigma_n = 10^{-42}$ cm$^2$. Starting from these rates, $\Gamma$, the number of events per unit time and mass of the detector is then computed as $R = \frac{\rho_\chi}{\mchi \nbar \mHe}\Gamma$, where $\rho_\chi = 0.3$~GeV/cm$^3$~\cite{Bovy:2012tw} is the local dark matter mass density. The comparison is reported in Figure~\ref{fig:comparison}.} 

{The two-phonon event signature is dominant in most of the mass range. However, for masses larger than roughly 500~keV, the ``cygnus-shaped'' three-phonon event signature becomes more relevant. Indeed, it is only for quite low masses that the two phonon events are dominated by back-to-back configurations. At higher masses, requiring such a configuration drastically reduces the rate.}


\section{Conclusion}

In this work we put forth the idea of looking for the process of emission of three phonons by a dark matter particle in superfluid \He. If the dark matter {is not too much lighter than $\mchi\simeq500$~keV}, this event can allow for triggering, vertex reconstruction, and directionality due to its peculiar kinematical configuration. This important information, which cannot be obtained from events involving one or two phonons, can be key towards an efficient background discrimination, and mass reconstruction.
{Moreover, if one enforces approximately back-to-back phonons for coincidence requirements, the three-phonon event considered here becomes dominant over the two-phonon one.
}

The dynamical matrix elements have been obtained from the relativistic EFT for superfluids, while the corresponding four-body phase space has been computed with both semi-analytical methods and fully numerical Monte Carlo techniques. The combination of these tools allows excellent control over this and many other observables, also confirming the EFT approach as a powerful theoretical framework, with possible applications to other materials or signatures, as well (see, e.g.,~\cite{Esposito:2020hwq}). {The four-phonon interaction vertex found here can be hard to obtain with standard condensed matter methods, and can hence be relevant for other studies on the dynamics of \He.}

The predicted rate for the process of interest is straightforward to obtain with the tools described here. However, its precise determination depends on the allowed angular and momentum resolutions for the outgoing phonons, which are ultimately given by the detector acceptance. 

Finally, as already noted, the three-phonon tree-level rates involving almost collinear final states are divergent. If one was interested in computing these configurations, the cancellation of these divergences, as well as their power counting within the EFT (see, e.g.,~\cite{Son:2005rv,Escobedo:2010uv,Dubovsky:2011sj,Berezhiani:2020umi}), should be  addressed.


\begin{acknowledgements}
We are grateful to G.~Cavoto, C.~Enss, L.~Gastaldo, J.~Jaeckel, A.~Nicolis and R.~Rattazzi for enlightening discussions, and especially to G.~Cuomo and A.~Monin for very fruitful exchanges on the soft divergences in Lorentz violating media. A.E. is supported by the Swiss National Science Foundation under contract 200020-169696, and through the National Center of Competence in Research SwissMAP. A.C. acknowledges support from the the Israel Science Foundation (Grant No. 1302/19), the US-Israeli BSF (Grant No. 2018236) and the German-Israeli GIF (Grant No. I-2524-303.7). A.C. acknowledges hospitality from the MPP of Munich.
\end{acknowledgements}


\appendix

\section{Matrix elements} \label{app:M}

Here we present the complete expressions for the matrix elements associated with the diagrams in Figure~\ref{fig:3phonon_diags}. Note that diagram $(b)$ and $(d)$ should appear three times, one for each permutation of the phonon eternal momenta. Here we report only one of them. The matrix elements are
\begin{widetext}
    \begin{subequations}
    \begin{align}
        \begin{split}
        \mathcal{M}_a & = \frac{G_{\chi } m_{\chi } \alpha \, \omega}{\omega^2-\cs^2q^2+ i \kappa \cs q^6} \bigg\{ \lambda_1 \left( \sprod{1}{2} \, \bm q_3\cdot \bm q + \sprod{1}{3} \, \bm q_2 \cdot \bm q + \bm q_1 \cdot \bm q \, \sprod{2}{3} \right) \\
        & \quad \lambda_2 \left( \omega_1 \, \omega_2 \,\bm q_3\cdot \bm q + \omega_1 \, \omega_3 \, \bm q_2 \cdot \bm q + \omega_1 \, \omega \, \sprod{2}{3} + \omega_2 \, \omega_3 \, \bm q_1 \cdot \bm q + \omega_2 \, \omega \, \sprod{1}{3} + \omega_3 \, \omega\, \sprod{1}{2} \right) + \lambda_3 \,\omega_1 \, \omega_2 \, \omega_3 \, \omega \bigg\} \,, 
        \end{split} \\
        \mathcal{M}_b & = - \frac{G_\chi \mchi \alpha \,\omega \left[g_1 \big( \omega_1 \, q_2^2 + \omega_2 \, q_1^2 + 2 \left( \omega_1 + \omega_2 \right) \sprod{1}{2} \big) + g_2 \, \omega_1 \, \omega_2 \left( \omega_1 + \omega_2 \right) \right]}{\left( \omega^2-\cs^2q^2+ i \kappa \cs q^6 \right) \left[ \left( \omega_1 + \omega_2 \right)^2 - \cs^2 \left(\bm q_1 + \bm q_2 \right)^2 + i \kappa \cs \left( \bm q_1 + \bm q_2 \right)^6 \right]} \\
        & \quad \times \Big[g_1\big( \omega_1\, \bm q_3 \cdot \bm q + \omega_2\, \bm q_3 \cdot \bm q + \omega_3 \left( \bm q_1 + \bm q_2 \right) \cdot \bm q + \omega \left(\bm q_1 + \bm q_2 \right) \cdot \bm q_3 \big)+ g_2\,  \omega \, (\omega_1+ \omega_2)\, \omega_3\Big] \,, \notag \\
        \mathcal{M}_c & = G_\chi \mchi \big[ \gamma_1 \left( \omega_1\, \sprod{2}{3} + \omega_2\, \sprod{1}{3} + \omega_3\, \sprod{1}{2} \right) + \gamma_2\, \omega_1\,\omega_2\,\omega_3 \big]\,, \\
        \begin{split}
        \mathcal{M}_d & = \frac{G_\chi \mchi \left[g_1 \big( \omega_1 \, q_2^2 + \omega_2 \, q_1^2 + 2 \left( \omega_1 + \omega_2 \right) \sprod{1}{2} \big) + g_2 \, \omega_1 \, \omega_2 \left( \omega_1 + \omega_2 \right) \right]}{\left( \omega_1 + \omega_2 \right)^2 - \cs^2 \left(\bm q_1 + \bm q_2 \right)^2 + i \kappa \cs \left( \bm q_1 + \bm q_2 \right)^6} \\
        & \quad \times \big[ \beta_1 \left( \bm q_1 + \bm q_2 \right) \cdot \bm q_3 + \beta_2 \,(\omega_1 + \omega_2)\, \omega_3  \big] \,,  \end{split}
    \end{align}
    \end{subequations}
\end{widetext}
where $\kappa \equiv \frac{\left( 3g_1+g_2\cs^2 \right)^2}{960\pi\cs^2} = \frac{\left(1 + \mHe \nbar \cs \frac{d\cs}{dP}\right)^2}{240\pi \mHe \nbar}$ characterizes the phonon width, while $\omega$ and $\bm q$ are, respectively, the total energy and momentum released by the dark matter.

When computed in the limit of small exchanged momentum, $\cs q \ll \omega$, the above matrix elements cancel pairwise---i.e. $\mathcal{M}_a + \mathcal{M}_c = O\left(q^2\right)$ and $\mathcal{M}_b + \mathcal{M}_d = O\left(q^2\right)$---, in order to respect the Ward identity following from the conservation of the superfluid particle number~\cite{Caputo:2019xum}. Despite the complicated structure of the matrix elements themselves, this pairwise cancellation can be easily understood just from the Feynman diagrams. As shown in~\cite{Caputo:2019cyg}, in the limit of small exchanged momentum one can integrate out the very off-shell phonon appearing in diagrams $(a)$ and $(b)$. This amounts to reducing its intermediate propagator to a contact interaction, hence precisely obtaining  diagrams $(c)$ and $(d)$, but with opposite sign. Interestingly, the derivative nature of the phonon coupling compensates for the nonlocality of the propagator, ultimately ensuring the locality of the effective interaction obtained by integrating out the gapless off-shell phonon.


\bibliographystyle{apsrev4-1}
\bibliography{biblio}

\end{document}